# Astro2020 Science White Paper

# Relativistic Jets in the Accretion & Collimation Zone: New Challenges Enabled by New Instruments

**Thematic Areas:** ☐ Planetary Systems    ☐ Star and Planet Formation
☒ Formation and Evolution of Compact Objects    ☒ Cosmology and Fundamental Physics
☐ Stars and Stellar Evolution    ☐ Resolved Stellar Populations and their Environments
☐ Galaxy Evolution    ☐ Multi-Messenger Astronomy and Astrophysics

**Principal Author:**
Name: Eric S. Perlman
Institution: Florida Institute of Technology
Email: eperlman@fit.edu
Phone: 321-674-7741

**Co-authors:** (names and institutions)
Mark Birkinshaw (University of Bristol), Matthias Kadler (Würzburg University), Serguei Komissarov (University of Leeds), Matthew Lister (Purdue University), David Meier (Caltech), Eileen Meyer (University of Maryland, Baltimore County), Masanori Nakamura (ASIAA), Kristina Nyland (NRAO), Christopher O'Dea (University of Manitoba), Diana Worrall (University of Bristol), Andrzej Zdziarski (Warsaw Observatory)

**Abstract** (optional):
Jets are a ubiquitous part of the accretion process, seen in a wide variety of objects ranging from active galaxies (AGN) to X-ray binary stars and even newly formed stars. AGN jets are accelerated by the supermassive black hole of their host galaxy by a coupling between the magnetic field and inflowing material. They are the source for many exciting phenomena and can profoundly influence the larger galaxy and surrounding cluster.

This White Paper points out what advances can be achieved in the field by new technologies, concentrating on the zone where jets are accelerated to relativistic speeds and collimated. The ngVLA and new space VLBI missions will give higher angular resolution, sensitivity and fidelity in the radio, penetrating this zone for additional objects and allowing us to resolve fundamental questions over the physics of jet acceleration and collimation. Interferometry in other bands would allow us to probe directly flaring components. We also emphasize the need for polarimetry, which is essential to revealing the role and configuration of magnetic fields.

Extragalactic jets were first observed a century ago, with Heber Curtis noting a "curious straight ray" in the galaxy M87. In active galactic nuclei (AGN), jets are likely created near the SMBH by a coupling between the magnetic field and accreting material (see Tchekhovskhoy 2015 for a review). The jet accelerates to nearly light speed in a distance that can range from as little as 1 pc to as much as hundreds of pc (Komissarov et al. 2007). Jets accelerate particles to extremely high energy, perhaps up to EeV energies. Jets constitute a major portion of the central object's energy budget, with a kinetic and radiative energy output that can approach or exceed Eddington levels. This allows them to influence profoundly the surrounding galaxy and cluster.

This White Paper concentrates on jet physics in the acceleration and collimation zone (ACZ, Blandford, Meier & Readhead 2018), where the jet emerges from the central accretion flow, and is collimated and accelerated to relativistic speeds. Our purpose is to emphasize what advances may be achieved by new technologies. Roughly speaking, this region covers from the SMBH's event horizon at out to the Bondi radius at ~$10^6$ $R_g$.

During the last several years, the jet has been resolved in the ACZ (Asada & Nakamura 2012, Nakamura & Asada 2013, Asada et al. 2014) for M87, one of the very nearest jets, and more recently for a handful of other objects. This has enabled critical advances to be made in jet physics. However, many questions remain in this region. These include the following:

1. Is the jet actually confined by the external medium, and if so how?
2. What is the nature of the jet in the ACZ, both in terms of particles and fields?
3. How quickly does the jet accelerate to speeds near c?
4. Are high-energy flares produced in the ACZ or elsewhere?
5. Does the jet recollimated beyond the ACZ, and if so, what physics are involved?
6. How does jet physics in the ACZ change as a function of jet power?

I.  **Imaging and Polarimetry in the ACZ and transition zones**

Within the ACZ, VLBA and GMVA observations to date indicate that most objects have an edge-brightened jet, with a width that increases smoothly with radius. M87's ACZ-scale jet (Fig 1) is mostly unpolarized, and shows only modest rotation measures (Hada et al. 2016, Algaba et al. 2016, Agudo et al. 2018, Walker et al. 2018). This indicates a low-density medium in the ACZ. *VLA* and *HST* images show very different morphology, with helical undulations (Owen et al. 1989) downstream of knot HST-1, 0.86" from the nucleus (Biretta et al. 1999). This and other evidence marks HST-1 as a possible *recollimation shock* (Stawarz et al. 2006, Asada et al. 2014).

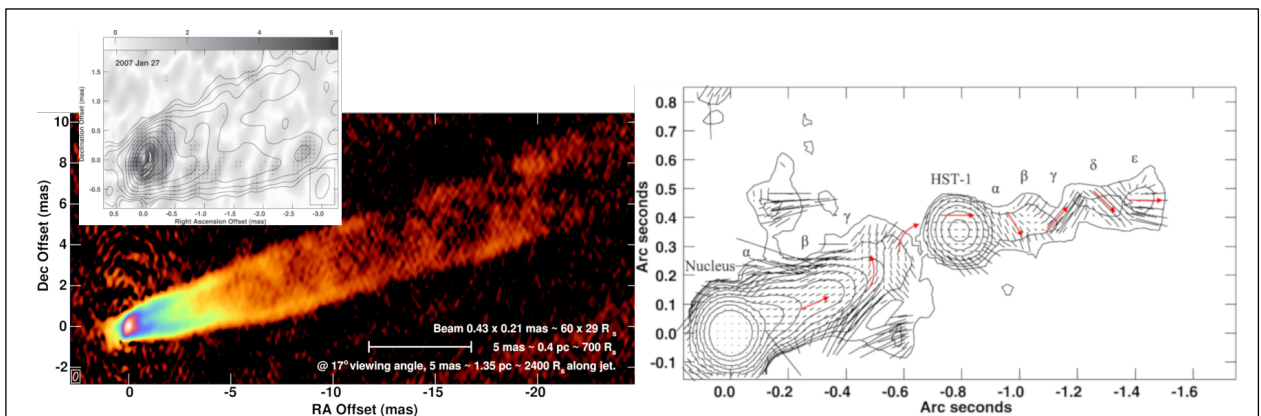

**Figure 1.** The inner jet of M87. The top two figures show the innermost 25 milliarcseconds, as seen with the VLBA at 43 GHz (Walker et al. 2018). The left image is total intensity, while its inset shows intensity in greyscale and contours with polarization vectors (electric field). The limb brightened nature of the jet on these scales is apparent. The JVLA image, at right, shows the larger-scale view of the ACZ plus the HST-1 region (Avachat et al. 2016). The contours represent total flux while the vectors represent fractional polarization (magnetic field). The helical undulations in the HST-1 region are striking.

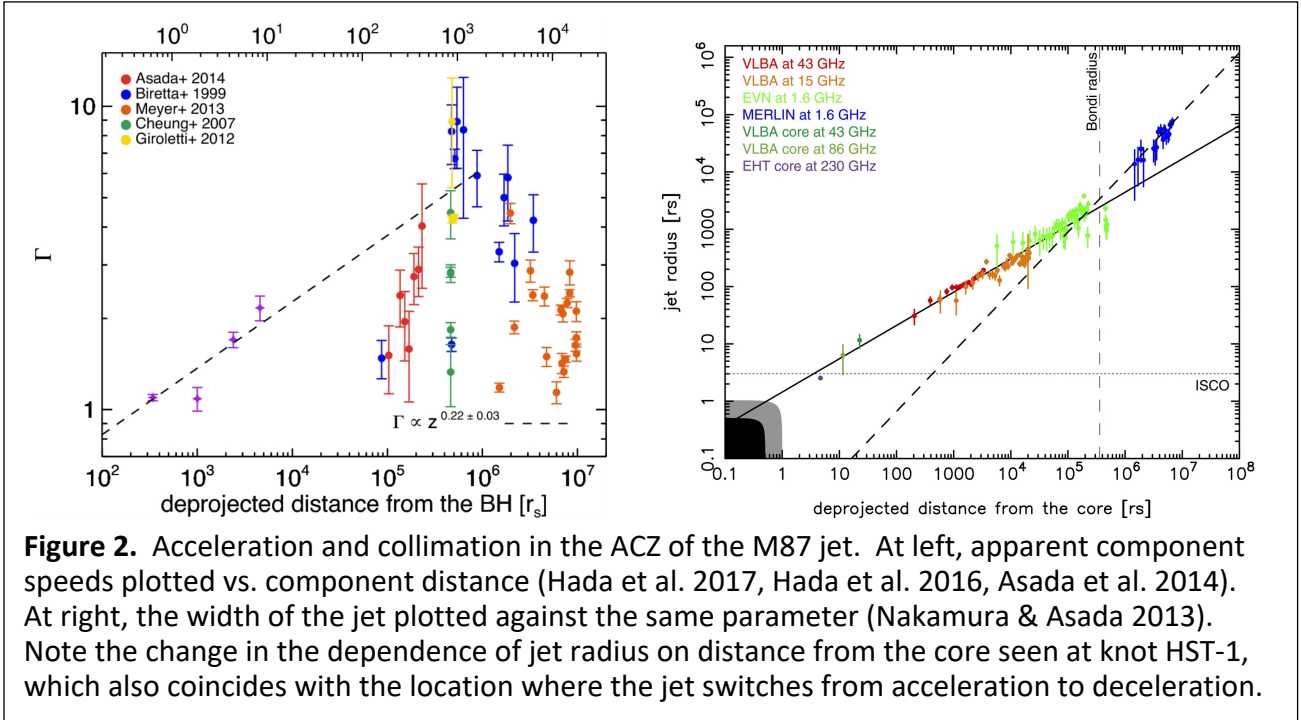

**Figure 2.** Acceleration and collimation in the ACZ of the M87 jet. At left, apparent component speeds plotted vs. component distance (Hada et al. 2017, Hada et al. 2016, Asada et al. 2014). At right, the width of the jet plotted against the same parameter (Nakamura & Asada 2013). Note the change in the dependence of jet radius on distance from the core seen at knot HST-1, which also coincides with the location where the jet switches from acceleration to deceleration.

Other jets present a somewhat different picture, and with more distant objects, the use of RadioAstron, giving baselines >1 $R_\oplus$, is helpful. Small Faraday rotations and smooth expansion are also observed in 3C 120 (Gomez et al. 2011) and NGC6251 (Tseng et al. 2016), indicating low-density media. However, the knots in 3C 120's ACZ are more highly polarized than those in M87. But the much larger rotation measure, nearly constant width and Faraday depolarization observed in 3C 84, combined with its nearly 90° bend (Kim et al. 2019, Giovannini et al. 2018), indicate a much higher-pressure cocoon or magnetized confining sheath. But more distant objects, such as BL Lac (z=0.069, Cohen et al. 2015) show only a stationary feature within the first milliarcsecond and cannot be studied within the ACZ. Overall, the small sample of jets studied in the ACZ make it hard to generalize regarding physics on this scale.

### II. The Recollimation feature and Kinematics in the ACZ

Here too the picture is complex but affected by the small number of objects with resolved ACZ regions. In some objects (e.g., M87), the jet's acceleration appears to continue for >100 pc. The recollimation feature in M87, known as knot HST-1, coincides with where the nature of the jet's pressure balance abruptly changes, as evidenced by the change from jet width $\propto r^{1.73}$ to $\propto r^{0.96}$ (Figure 2). HST-1 is also where the superluminal speeds seen show a switch from acceleration to deceleration. In some other jets (NGC 6251, 3C84), superluminal components have not been seen in the ACZ (Lister et al. 2016, 2019). But in 3C 120, which is more powerful (Figure 3, Lister et al. 2016, 2019), we see many superluminal components in the first 20 milliarcseconds, while the recollimation feature is 80 milliarcseconds from the core. Recollimation features are seen in other nearby blazar and quasar jets, for example BL Lac, which shows a stationary feature 0.26 milliarcsec from the core and superluminal components beyond that. In more distant objects, superluminal motion is seen (Lister et al. 2017, 2018), but the recollimation shock may be on scales too small to be observed (Marscher et al. 2008). The birth of superluminal components has been linked to gamma-ray flares (Casadio et al. 2015).

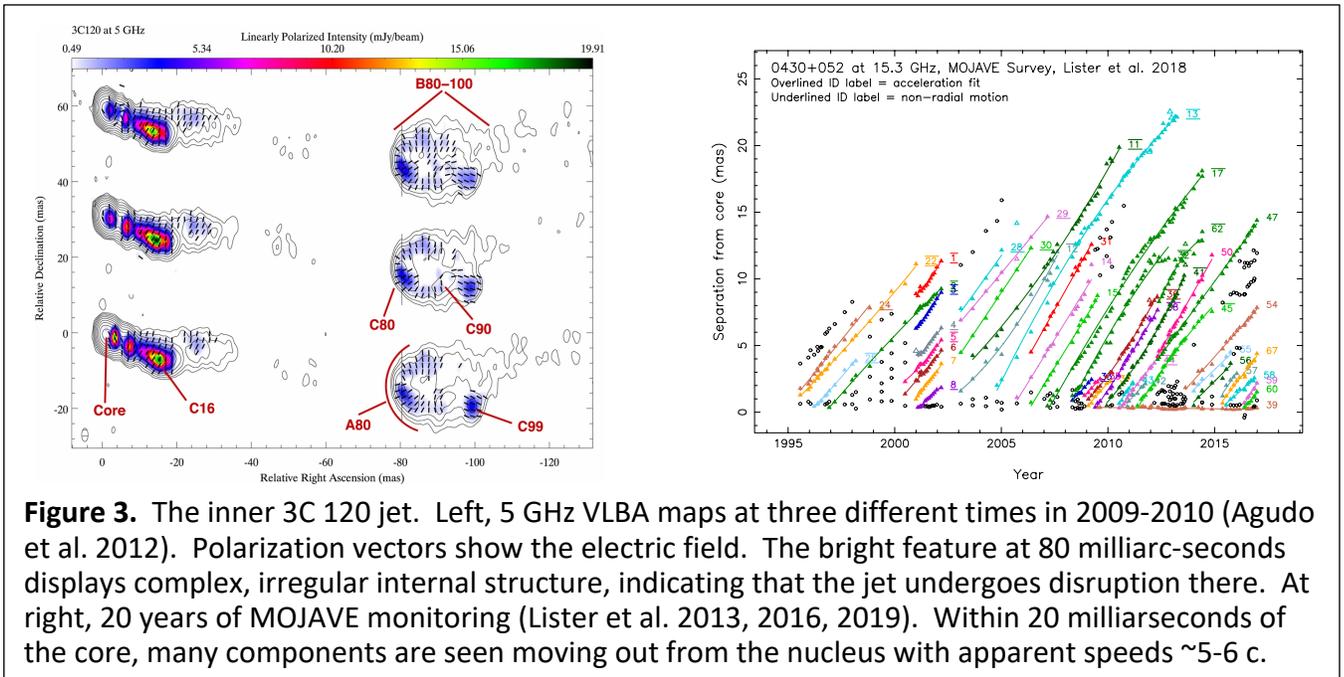

**Figure 3.** The inner 3C 120 jet. Left, 5 GHz VLBA maps at three different times in 2009-2010 (Agudo et al. 2012). Polarization vectors show the electric field. The bright feature at 80 milliarc-seconds displays complex, irregular internal structure, indicating that the jet undergoes disruption there. At right, 20 years of MOJAVE monitoring (Lister et al. 2013, 2016, 2019). Within 20 milliarseconds of the core, many components are seen moving out from the nucleus with apparent speeds ~5-6 c.

### III. Dynamical Balance and Nature of the Recollimation Shock

The dynamical balance inside the ACZ is not clear, and neither is the nature of the recollimation feature. One view (Komissarov & Falle 1997, Stawarz et al. 2006) is that the jet expands freely inside the ACZ and is pressure-confined beyond the ACZ. In this picture, the recollimation feature is a structural transition caused by the lack of pressure confinement, as well as a change in the balance between the relativistic centrifugal instability and the Kelvin-Helmholtz instability (Gorgouliatos & Komissarov 2018a, b). A second picture (Asada et al. 2014, Globus & Levinson 2016, Levinson & Globus 2017) holds that inside the ACZ the jet is thermally confined by a stratified ISM created by the outflow wind and accretion flow. In this picture, the jet is overpressurized at the recollimation shock and changes at that point to one that is freely expanding. The transformation of the jet's structure and magnetic field at this location is likely triggered by the 3D magnetic kink instability, which is excited by the break in the density profile. Modeling by Barniol Duran et al. (2017, Figure 4) suggests that recollimation may be triggered by a change in the density profile of the surrounding medium, where the inner jet's flow has ceased to push out a 'bubble' in the nuclear ISM. The location of stationary features and structural changes near the Bondi radius suggests a link between the recollimation and the accretion process (Asada & Nakamura 2012, Nakamura & Asada 2013).

### IV. Impact of New Observational Tools

All of the questions detailed at the beginning of this White Paper can benefit from new observational and theoretical tools, and in fact the small number of objects studied in the ACZ makes this critical. The building of the ngVLA and future radio astronomy missions will make it possible to achieve still higher angular resolutions. This will help to resolve more jets in the ACZ, making it possible to view the collimation and acceleration for many more objects and better elucidate the nature of the jet in this region. In this final section, we detail how additional observational tools will do this.

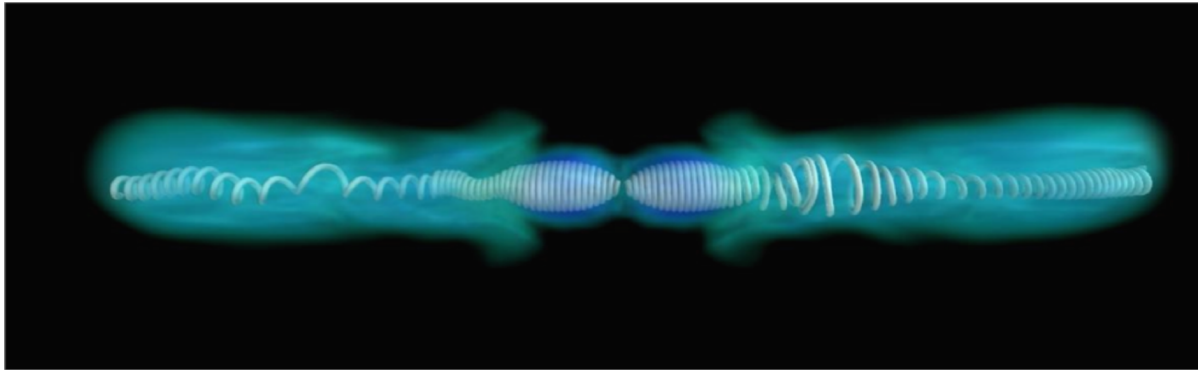

**Figure 4.** A 3D rendering of a jet simulation (Barniol Duran et al. 2017). The SMBH is located at the center. The white curve shows a representative magnetic field line, which is twisted as the SMBH rotates. The sheath "funnel" marks a break in the nuclear density profile, which causes the jets to recollimate and the toroidal magnetic field to build up and become unstable to the 3D magnetic kink instability. The bends and asymmetries of the magnetic field lines are tell-tale signs of the instability. The instability dissipates toroidal magnetic field into heat around the density break and leads to a less tightly wound magnetic field at large radii.

**Revealing the ACZ: a better understanding of jet acceleration and collimation.** Perhaps the most striking thing about the observations we detailed is how few objects have been observed in the ACZ. The only objects for which the ACZ is observationally accessible at present are very low-redshift, FR I objects that also have massive SMBH. The former requirement excludes almost all high-power, FR II jets. A dramatic illustration is shown in Figure 5 (Nakahara et al. 2018), which shows the comparison between observations to date for the M87, NGC 6251, NGC 4261 and Cygnus A jets. As can be seen, in Cygnus A we just penetrate the ACZ and do not presently see a recollimation feature, meaning that we cannot study in detail the physical processes involved in its acceleration and confinement. The story is similar for 3C 273 (Akiyama et al. 2018), where the larger black hole mass compensates for the greater distance.

To remedy this, we need higher angular resolution. This will allow us to observe the ACZ in more objects, and will also reveal new recollimation features. We may also be able to observe the ejection and acceleration of new superluminal components that may accompany energetic variability events. It will also help resolve the issue of the dynamical balance within the ACZ and the nature of recollimation features and link them to jet acceleration. In addition, we need to probe any ACZ-scale differences between the low-power, FR I jets, and their rarer but higher-power FR II cousins. Finally, we could also probe the initial collimation zone of the jet, only now being observed for a few sources by the observations at 1.3mm and 3mm EHT. This would be accomplished by additional space VLBI missions (in addition to RadioAstron's replacement) and 1.3mm (EHT)/3mm GMVA+GLT observations. Tantalizing hints of FR I/FR II differences are seen in existing data: *e.g.*, much higher Faraday rotations in 3C 273 than in M87 (Akiyama et al. 2018), which could be a result of a much larger accretion flow, consistent with its much higher luminosity, and deviations from conical flow in blazars and narrow-line Seyfert I galaxies (Algaba et al. 2017, Pushkarev et al. 2017, Hada et al. 2018).

**Increased Fidelity: Multi-wavelength Studies in the Radio, Understanding the Jet Spectrum.**
A second need is for better coverage of intermediate scales as well as increased fidelity,

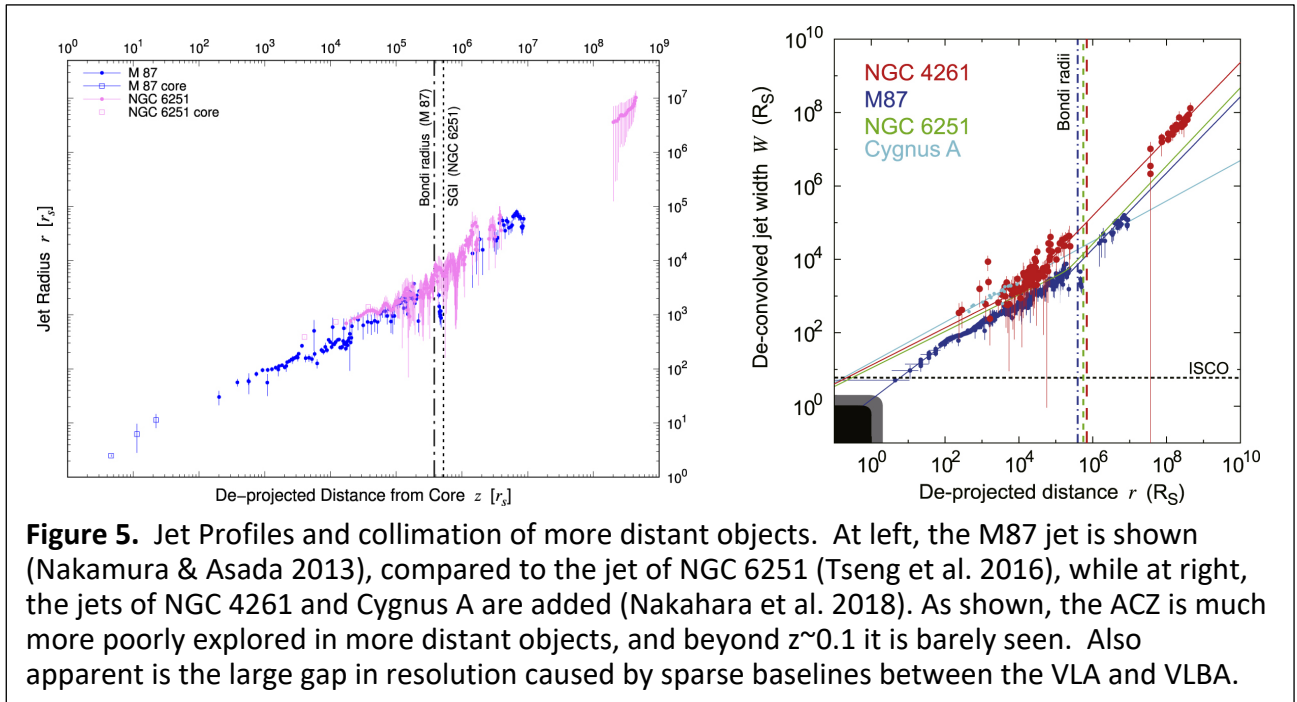

**Figure 5.** Jet Profiles and collimation of more distant objects. At left, the M87 jet is shown (Nakamura & Asada 2013), compared to the jet of NGC 6251 (Tseng et al. 2016), while at right, the jets of NGC 4261 and Cygnus A are added (Nakahara et al. 2018). As shown, the ACZ is much more poorly explored in more distant objects, and beyond z~0.1 it is barely seen. Also apparent is the large gap in resolution caused by sparse baselines between the VLA and VLBA.

sensitivity and spectral coverage. Current data (Fig 1, Fig 5) require the use of MERLIN, EVN and VLBA at multiple frequencies. The use of multiple frequencies introduces complications, as complex spectral structure likely exists. It would be much better to have continuously high sensitivity with a densely sampled range of baseline lengths from hundreds of meters all the way up to 10,000+ km at all wavelengths, which only the ngVLA can offer. Nyland et al. (2018) present a powerful argument for what the ngVLA could accomplish. Not only could we see the differences between many different kinds of jets, but in addition, its high fidelity over a wide range of frequencies would allow unprecedented spectral studies, allowing us to not only see particle acceleration and aging in jets that we currently know a lot about, but also study physics in much younger objects. The SKA, LWA, EHT and ALMA would allow us to extend this to higher and lower frequencies, helping us to better understand spectral structure in the ACZ.

**Optical/Near-IR Interferometry: Firmly linking ACZ Jets and High-Energy Flares.** A third need here is to image the innermost components of jets in other wavebands. This will be possible in the near-infrared and optical within the next decade, thanks to a variety of instruments and techniques. These include the TESS probe in the optical as well as with ground-based interferometers including the VLT + GRAVITY and (potentially) the Keck interferometer. X-ray interferometry is also under study. The exciting results on resolving the accretion disk of 3C 273 (Sturm et al. 2018) make clear the potential for breaking fundamental new ground. By imaging the ACZ-scale jet in the near-IR, optical or X-ray, we could directly link gamma-ray variability to the ejection of new components. Currently this has been done for precisely one object: M87, and only during the 2005 HST-1 flare (Harris et al. 2009; Perlman et al. 2011; Abramowski et al. 2012), where the component was both far removed from the core and the object as close as possible. Without much higher angular resolution in the optical, current work (e.g., Rani et al. 2018, Sasada et al. 2018, Raiteri et al. 2017), is attempting to draw links across particle populations that differ by at least two if not three orders of magnitude in particle energy to the radio, far below the particle energies that are Comptonized up to GeV and higher energies, where we do not see the new component until weeks if not months after the gamma-ray flare.